\documentclass[aps,preprint]{revtex4}
\usepackage{graphicx}
\usepackage{amssymb}

\makeatletter

\begin{document}
\global\long\def\met{\not{\!{\rm E}}_{T}}

\begin{flushright}
ANL-HEP-PR-10-8, EFI-10-4, NUHEP-TH/10-02 
\end{flushright}

\title{Higgs Boson Search Sensitivity in the $H \rightarrow WW$ Dilepton
Decay Mode at $\sqrt s = 7$ and $10$~TeV}
\author{ Edmond L. Berger$^{a}$}
\email{berger@anl.gov}
\author{Qing-Hong Cao$^{a,b}$}
\email{caoq@hep.anl.gov}
\author{C.~B.~Jackson$^{a}$}
\email{jackson@hep.anl.gov}
\author{Tao Liu$^b$}
\email{taoliu@theory.uchicago.edu}
\author{Gabe Shaughnessy$^{a,c}$}
\email{g-shaughnessy@northwestern.edu} 
\affiliation{
\mbox{$^a$High Energy Physics Division, Argonne National Laboratory, Argonne, IL
60439}  \\ 
\mbox{$^b$Enrico Fermi Institute, University of Chicago, Chicago, IL 60637}  \\
\mbox{$^c$Department of Physics and Astronomy, Northwestern University, Evanston,
IL 60208}
} 
\begin{abstract}
Prospects for discovery of the standard model Higgs boson are examined at 
center of mass energies of $7$ and $10$ TeV at the CERN Large Hadron Collider. 
We perform a simulation of the signal and principal backgrounds for Higgs boson 
production and decay in the $W^+ W^-$ dilepton mode, finding good agreement 
with the ATLAS and CMS collaboration estimates of signal significance at 14 TeV 
for Higgs boson masses near $m_H = 160$~GeV.  At the lower energy of $7$~TeV, 
using the same analysis cuts as these collaborations, we compute expected signal 
sensitivities of about $2$ standard deviations ($\sigma$'s) at $m_H = 160$~GeV in 
the ATLAS case, and about 3.6~$\sigma$ in the CMS case for $1$~fb$^{-1}$ of 
integrated luminosity.   Integrated luminosities of 8~$\rm{fb}^{-1}$ and 
3~$\rm{fb}^{-1}$ are needed in the ATLAS case at $7$ and $10$~TeV, respectively, 
for $5~\sigma$ level discovery.  In the CMS case, the numbers are 2~$\rm{fb}^{-1}$ 
and 1~$\rm{fb}^{-1}$ at $7$ and $10$~TeV.  Our different stated expectations for the 
two experiments arise from the more restrictive analysis cuts in the CMS case.  
Recast as exclusion limits, our results show that with $1~{\rm fb}^{-1}$ of integrated luminosity 
at 7~TeV, the LHC may be able to exclude $m_H$ values in the range 160 to 
180~GeV provided no signal is seen.    

\end{abstract}

\maketitle

\section{introduction}

Discovery of the Higgs boson of electroweak symmetry breaking is a prime goal of 
experimental investigations at the CERN Large Hadron Collider (LHC).    Considerable 
efforts have been made by the ATLAS and CMS collaborations to simulate the 
production and decay of the standard model Higgs boson at the LHC's 
center-of-mass (cm) design energy of $14$~TeV~\cite{Aad:2009wy, CMS:PPR}, and 
estimates have 
been made of the integrated luminosity required to observe the Higgs boson and to 
measure its properties.   The recent decision to begin operation of the LHC at the 
reduced cm energy of $7$~TeV (and possibly $10$~TeV) \cite{chamonix:2010} motivates an examination of the 
discovery potential at these lower energies.   In this paper we present our 
estimates of the potential for finding the Higgs boson through its $W^+ W^-$ decay 
mode at $7\,{\rm TeV}$ and $10\,{\rm TeV}$.  

At both the Fermilab Tevatron and the LHC the largest channel for production
of the Higgs boson is gluon fusion, $g g \rightarrow H X$, with the $ggH$ coupling 
arising via a top quark loop (cf. Refs.~\cite{Gunion:1989we,Djouadi:2005gi}).  The relative dominance of production and decay modes  
depends on the Higgs boson mass.  A Higgs boson with mass $m_{H}\gtrsim135$
GeV decays predominantly to $W$ boson pairs with one of the $W$'s
potentially off mass-shell.  In this mass range inclusive production through gluon 
fusion is dominant at the LHC,  with the best sensitivity
occurring around $m_{H}\approx160-170$ GeV, where the $WW$ decay
mode is fully open. 

In this paper we limit ourselves to the gluon fusion production process and to the 
$WW$ decay mode, focusing on observation in the dilepton channel in which 
$H \rightarrow W^+ W^- \rightarrow l^+ l^-$ plus missing energy. 
We present a simulation of the signal and of the salient continuum $W^+ W^-$, 
$t \bar{t} X$, and $W + \rm{jets}$  backgrounds, applying the same analysis cuts used  
by the the ATLAS and CMS collaborations at $14$~TeV~\cite{Aad:2009wy, CMS:PPR}.   
In the ATLAS case, we conclude that integrated luminosities of 8~$\rm{fb}^{-1}$ and 
3~$\rm{fb}^{-1}$ are needed at $7$ and $10$~TeV, respectively, for $5\sigma$ level 
discovery of a standard model Higgs boson of mass $m_H = 160$ GeV in this 
production and decay mode.  In the CMS case, the numbers are 2~$\rm{fb}^{-1}$ and 
1~$\rm{fb}^{-1}$ at $7$ and $10$~TeV, respectively, for $5\sigma$ level 
discovery.   Optimization of the analysis cuts for 
$7$ and $10$~TeV might reduce the required luminosities.  The different analysis 
cuts explains the different required luminosities for the two experiments.   Larger 
samples would be needed for masses as low as $m_H=140$ GeV or above 
$m_H=180$ GeV.

There are important production channels in addition to $g g \rightarrow H$.    
These include production of the
Higgs boson with an associated $W$ boson, $Z$ boson, or top quark
pair, as well as production of the Higgs boson through vector-boson
or bottom-quark fusion.   A light Higgs boson ($m_{H}\lesssim135$
GeV) decays predominantly to bottom quark pairs. In this case the
inclusive Higgs boson signal is difficult to pick out from the large QCD 
$b\bar{b}$ background.    Other important Higgs boson decay
modes, both at the Tevatron and the LHC, are $ZZ$ for high mass Higgs
bosons, and tau meson pairs and photon pairs for low mass Higgs bosons.  
These and other modes were used in the recent combined fit of Tevatron 
data to exclude the mass range of $163\,{\rm GeV}<m_{H}<166\,{\rm {GeV}}$
at 95\% C.~L.~\cite{Phenomena:2009pt}.  

According to current expectations, the LHC will operate at 7 TeV for a couple of years 
or until it accumulates an integrated luminosity
 of $1\,{\rm fb}^{-1}$ \cite{chamonix:2010}.   With low luminosity it might be difficult
to observe a light Higgs boson owing to the small branching fraction to 
$\gamma\gamma$ , or a heavy Higgs boson in the $ZZ$ channel
because this decay  mode suffers from the small decay branching ratio of 
$Z\to\ell^{+}\ell^{-}$($\ell^{\pm}$
denotes charged leptons).    Based on such considerations,  we focus first on the 
leading decay 
channel $H\to WW$ throughout this work.   We intend to address the $ZZ$ case 
and other production and decay modes at a later date.   

The remainder of the paper is organized as follows. In Section~\ref{sect:xsec} we  present 
next-to-leading order (NLO) calculations of the production cross section of the 
Higgs boson and the principal backgrounds at the LHC for $7$, $10$, and $14$~TeV.   
We then describe our method for simulating the signal and the background processes, taking into 
account the different lepton momentum requirements and kinematic cuts used by the 
ATLAS and CMS collaborations.    As shown in Section~\ref{sec:Comparison}, our determinations of signal and background acceptances at $14$~TeV are in good agreement with those reported 
in the 2008 ATLAS Physics Performance Report (PPR)~\cite{Aad:2009wy} and, except in the 
$t \bar{t} X$ case,  also with the 2007 CMS PPR~\cite{CMS:PPR}.  We examine possible 
interpretations of this one disagreement.   In Section~\ref{sect:sensitivity} we use our 
acceptances and the NLO cross sections at lower LHC  energies to determine the discovery 
potential at $7$ and $14$~TeV.   A brief Section~\ref{sect:exclude} addresses Higgs boson 
exclusion limits.  We state our conclusions in Section~\ref{sect:conclusions}.  

\section{Cross sections and detection efficiencies}
\label{sect:xsec}

For the $H \to WW$ channel, ATLAS \cite{Aad:2009wy} and CMS \cite{CMS:PPR} present 
detailed studies of the signal and backgrounds at a cm energy of 14 TeV.  The collider signature of the signal events is characterized by two oppositely-charged leptons plus large missing energy originating from two invisible neutrinos.  The background processes include  
$t\bar{t} X$, $WW$, $WZ$, $ZZ$, $Z$+jets, and $W$+jets.  Isolated leptons from heavy-flavor pair production and semi-leptonic decay are also generally important~\cite{Sullivan:2006hb,Sullivan:2008ki}.  After suitable cuts~\cite{Sullivan:2006hb,Sullivan:2008ki}, however, the background for $m_H > 140$~GeV is dominated by 
$t\bar{t}$ production and continuum $W^{+}W^{-}$ pair production.   In the following, we focus on these two backgrounds, but we also examine the potential role of $W$+jets  which the ATLAS 
simulations suggest could be large, albeit with large uncertainties \cite{Aad:2009wy}.     In this 
work, we represent the $W$+jets contribution by $W + c$ production, with $c \rightarrow l X$.   

The ATLAS and CMS studies include sophisticated simulations of both signal and background event rates and also simulations of the detector response (e.g., lepton triggers and jet vetoes along with their associated efficiencies).   In this paper, we follow a more simplified approach.   An important check 
of our method is a comparison of our results for calculated acceptances and signal significance at 14 TeV with those of ATLAS and CMS.  In this section, we first present our NLO calculations of the inclusive cross sections for the signal and principal backgrounds, and then we outline the algorithm used to generate both signal and background events for the comparison at 14 TeV and for our predictions at 7 and 10 TeV.

\subsection{Next-to-leading Order Cross Sections}
\label{subsec:NLOcrosssections}

The Higgs boson production cross section in gluon-gluon scattering has been calculated
at leading order, next-to-leading order (NLO)~\cite{Dawson:1990zj}
and next-to-next-to-leading order (NNLO)~\cite{Harlander:2002wh,Anastasiou:2002yz,Ravindran:2003um}
in the infinite-top-quark-mass limit, and at LO and NLO~\cite{Djouadi:1991tka,Graudenz:1992pv}
with full top quark mass dependence. In addition to the QCD corrections, the NLO electroweak (EW)
corrections have also been considered in the infinite-top-quark-mass limit~\cite{Djouadi:1994ge},
and more complete calculations have been performed by including light quark
and top quark effects~\cite{Aglietti:2004nj,Degrassi:2004mx}. Recently, the effects of the combined QCD and EW
corrections were analyzed~\cite{Actis:2008ug,Anastasiou:2008tj}.

We use the MCFM~\cite{MCFM} code to compute the NLO inclusive cross sections for the signal 
and for background processes. The renormalization and factorization scales are set to $m_H$ for 
the $gg\to H$ signal,  to $m_t$ for $t \bar{t} X$, to $2 m_W$ for the $WW$ continuum and to $m_W$ for the $Wc X$ backgrounds.   
All cross sections are computed with the CTEQ6.6M parton distribution function (PDF) package~\cite{Nadolsky:2008zw}.   
In Table~\ref{tab:nlo-xsec}, we present the NLO signal cross sections for several values of the Higgs boson mass ($m_{H}=140,\,160,\,180$ and $200$ GeV) and for the major backgrounds.  
%
%
%
%

\begin{table}
\caption{Inclusive NLO cross sections (pb) for Higgs boson production and the $t\bar{t}$,
$WW$ and $Wc$ (both $W^+ c$ and $W^- c$) backgrounds.  \label{tab:nlo-xsec}}

\begin{tabular}{c|c|c|c|c|c|c|c}
\hline 
process &  & \multicolumn{2}{c|}{14 TeV} & \multicolumn{2}{c|}{10 TeV} & \multicolumn{2}{c}{7 TeV}\tabularnewline
\cline{3-8} 
 &  & LO & NLO & LO & NLO & LO & NLO\tabularnewline
\hline
\hline 
 & $m_{H}=140\,{\rm GeV}$ &\,\,11.3\,\, & \,\,25.7\,\, & \,\,6.33\,\, & \,\,14.3\,\, & \,\,3.25\,\, & \,\,7.36\,\, \tabularnewline
\cline{2-8} 
$gg\to H$ & $m_{H}=160\,{\rm GeV}$ & 9.16 & 20.6 & 5.01 & 11.2 & 2.51 & 5.62\tabularnewline
\cline{2-8} 
 & $m_{H}=180\,{\rm GeV}$ & 7.59 & 16.9 & 4.07 & 9.04 & 1.99 & 4.42\tabularnewline
\cline{2-8} 
 & $m_{H}=200\,{\rm GeV}$ & 6.44 & 14.2 & 3.39 &  7.47 & 1.62 & 3.57\tabularnewline
\hline
\hline 
$t\bar{t}$  & $m_{t}=175\,{\rm GeV}$ & 544 & 807 & 257 & 374 & 112 & 158\tabularnewline
$WW$ & $m_{W}=80.3\,{\rm GeV}$ & 83.1 & 115 & 52.8 & 71.9 & 31.5 & 42.5\tabularnewline
$Wc$ & $m_{W}=80.3\,{\rm GeV}$ & 6350 & 8450 & 3710 & 5000 & 2030 & 2720\tabularnewline
\hline
\end{tabular}
\end{table}

\begin{figure}
\includegraphics[scale=0.7,clip]{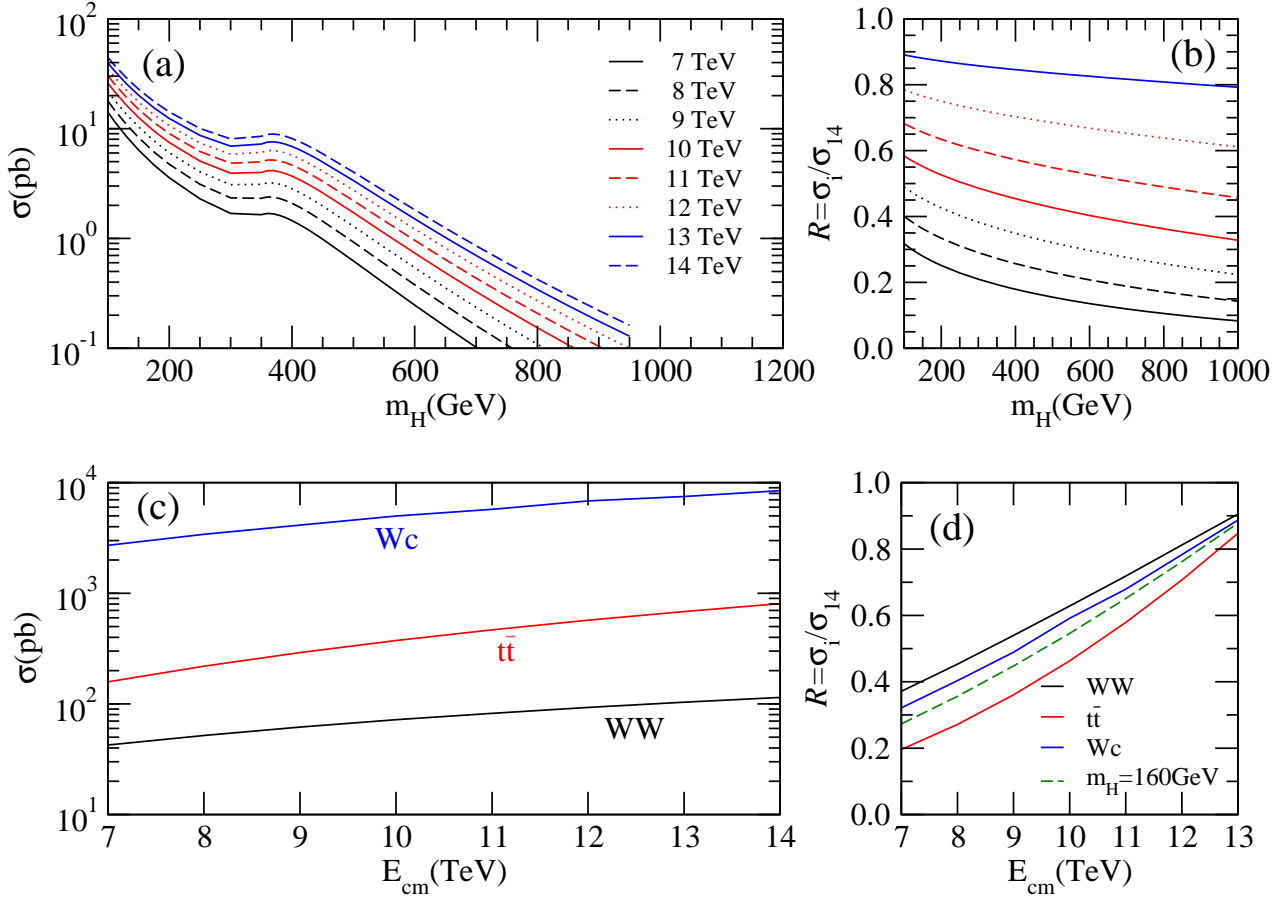}
\caption{(a) Total cross section for $gg \to H$ as a function of the Higgs boson mass for several different cm energies.  (b) Ratio of the cross sections for $gg \to H$ at different cm energies to the cross section at 14 TeV.  (c) Total cross sections for the backgrounds ($W^+W^-$, $t\bar{t}$ and $Wc$) as a function of the cm energy.  (d) Ratio of the background cross sections at different cm energies to the cross sections at 14 TeV. 
For comparison we also plot the signal ratio curve for  $m_H=160~\rm{GeV}$.
\label{fig:xsec-signal-bkgd}}
\end{figure}

In Fig.~\ref{fig:xsec-signal-bkgd}(a), we show the total cross section for the signal as a function of the Higgs boson mass for several cm energies.   Involving two gluons in the initial state, the signal cross section drops rapidly with decreasing cm energy as can be easily understood from a consideration of the behavior of the gluon PDF. The Higgs boson mass provides a natural choice of the physics scale 
resulting in an effective parton Bjorken-$x$, $\langle x \rangle \simeq m_H / \sqrt{s}$. 
For a given Higgs boson mass, lowering the cm energy increases the effective $x$.  Numerically,  
$\langle x \rangle \simeq m_H / \sqrt{s} \sim 0.01$ for a 140 GeV Higgs boson at 14 TeV and 
$\langle x \rangle \simeq m_H / \sqrt{s} \sim 0.02$ at 7 TeV. Since the gluon PDF drops rapidly with 
$x$, the signal cross section also decreases rapidly as $\sqrt s$ is decreased, 
as shown in Fig.~\ref{fig:xsec-signal-bkgd}(a).   Correspondingly, increasing the Higgs boson mass
at fixed $\sqrt s$ will force the effective $x$ to be larger, suppressing the cross section; see the 
broad band in the range $400~{\rm GeV} < m_H < 1000~{\rm GeV}$. 
To make the point clear we plot in Fig.~\ref{fig:xsec-signal-bkgd}(b)
the ratio $R$($=\sigma_{i}/\sigma_{14}$), defined as the ratio of the Higgs boson 
production cross section at a particular cm energy to the cross section at 14 TeV.
We observe that the cross section is reduced by a factor of 2 to 2.5 at 10 TeV
and by a factor of 3 to 8 at 7 TeV for a Higgs boson in the mass range of 100~GeV to 600~GeV.

The cross sections and the ratio $R$ for the backgrounds are shown as a function 
of cm energy in Fig.~\ref{fig:xsec-signal-bkgd}(c) and (d).
The $t\bar{t}$ background is produced mainly from the gluon-gluon initial state at  14~TeV 
(90\% from the gluon-gluon initial state and 10\% from the quark-antiquark initial state).
The $gg$ initial state remains the leading contributor at 7~TeV (about 80\%); see
Fig.~(3) in Ref.~\cite{Berger:2009qy}.  Therefore, lowering the cm energy 
decreases the $t\bar{t}$ background about as much as the Higgs signal; for example, 
$R(H) \approx R(t\bar{t}) \simeq 0.45$ and $0.20$ at 10~TeV and 7~TeV 
for $m_H = 2 m_t = 350~{\rm GeV}$. 
On the other hand, the $WW$ continuum background originates from the valence-quark and sea-quark
initial state, and it decreases less than the signal and the $t\bar{t}$ background; see
Fig.~\ref{fig:xsec-signal-bkgd}(c) and (d).   This difference is crucial for the Higgs boson search.  
Since the $WW$ continuum is the major background, the fact that the signal is suppressed 
more than the background at lower cm energy means that more integrated luminosity is needed 
to restore the same significance for Higgs boson discovery as at 14~TeV.  The $Wc$ contribution is produced dominantly by $gs\to Wc$ with a hard scale of roughly $m_W$ which leads to a typical parton-$x$ where PDF suppression is not strong.

\subsection{Generation of Event Samples}

The Higgs boson signal at the LHC consists of the leading-order (LO) process 
$pp \to H \to WW$ along with higher-order corrections from initial-state radiation which can produce multiple jets.
In order to simulate these effects, one may use event generators which include parton showering (such as PYTHIA \cite{Sjostrand:2007gs} or HERWIG \cite{Corcella:2002jc}) to generate event samples for both signal and backgrounds.  In some cases, next-to-leading order (NLO) event generators which correctly account for NLO QCD effects and initial-state radiation are available (e.g., MC@NLO \cite{Frixione:2002ik,Frixione:2006gn} and POWHEG \cite{Frixione:2007vw}).
In most cases a $p_T$-dependent $K$ factor obtained from a code such as MCFM is applied to the PYTHIA events in order to reweight the sample \cite{Aad:2009wy, CMS:PPR}. 

In this work, we adopt a slightly simplified approach to model total event rates and parton showering effects.  We generate the signal events with MadGraph/MadEvent \cite{Maltoni:2002qb}, while we use ALPGEN  \cite{Mangano:2002ea} to generate the background events.  This is done in an attempt to streamline the analysis since the matrix-elements-squared for the background processes are hard-coded in ALPGEN and, thus, are much more compact than those produced with MadGraph.  However, for the signal processes, ALPGEN does not currently include spin-correlation effects among the leptons from $H \to W^+ W^- \to \ell^+ \nu \ell^- \bar{\nu}$.  We choose to use MadGraph/MadEvent to produce events for the signal since spin-correlation effects are crucial for light Higgs boson searches \cite{Anastasiou:2009bt, Anastasiou:2008ik}.  At leading order, the signal process is:
\begin{equation}
p p \to H \to W^+ W^- \to \ell^+ \nu \ell^- \bar{\nu} \,,
\label{eq:sigLO}
\end{equation}
while the irreducible background process is:
\begin{equation}
p p \to W^+ W^- \to \ell^+ \nu \ell^- \bar{\nu}\,,
\label{eq:bkgWWLO}
\end{equation}
and the two reducible background processes are:
\begin{eqnarray}
p p &\to& t\bar{t} \to W^+ W^- b \bar{b} \to \ell^+ \nu \ell^- \bar{\nu} b \bar{b}  \,, \\
p p &\to& W  c  \to \ell \nu c\, ,
\label{eq:bkgttLO}
\end{eqnarray}
where the renormalization and factorization scales in all processes are chosen in accordance with the values mentioned in the previous section.

The reducible background from $t\bar t$ occurs when both $b$-quarks are not tagged as jets.  The background of $Wc$ masks the signal topology when the $c$-jet decays semi-leptonically to a tagged lepton. We assume an isolated lepton probability of $\epsilon_{c\to \ell} = 0.5$\% when the $c$-quark is in the region $p_T>20$ GeV and $|\eta| < 1$~\cite{Sullivan:2006hb,Sullivan:2008ki}.  
The momentum imparted to the isolated lepton is roughly 90-95\% of the parent heavy flavor quark~\cite{Sullivan:2006hb,Sullivan:2008ki}.  The ATLAS collaboration includes the possibility of light jets faking isolated leptons near the level of few$\times10^{-5}$.  However, owing to the gluon PDF, the $gs\to Wc$ subprocess is enhanced, and the subsequent dilepton rate dominates the total $W^\pm+$jets rate.  Therefore, we only consider $W^\pm c$ events and do not include light jet fakes.

To mimic the effects of initial state radiation and parton showering, we include the possibility of additional jets in the final state.  For example, in addition to the LO signal process of Eq.~(\ref{eq:sigLO}), we also generate events for:
\begin{equation}
p p \to H + nj \to W^+ W^- + nj \to \ell^+ \nu \ell^- \bar{\nu} + nj\,,
\label{eq:sig+jets}
\end{equation}
where $n$ = 1 or 2 and $j$ denotes a light jet.  The events from these processes are then combined 
with those of the LO process, and the sum is normalized to the total NLO event rate (we use 
inclusive K factors derived from the cross sections shown in Table~\ref{tab:nlo-xsec}) to produce an effectively ``showered'' final state.  Similarly, for the backgrounds, we add to the processes of Eqs.~(\ref{eq:bkgWWLO}-\ref{eq:bkgttLO}) events from:
\begin{eqnarray}
p p &\to& W^+ W^- + nj \to \ell^+ \nu \ell^- \bar{\nu} + nj\,,\\
p p &\to& W  c  +nj \to \ell \nu c + nj\,,
\label{eq:bkgWWLOnj}
\end{eqnarray}
and:
\begin{equation}
p p \to t\bar{t} + nj \to W^+ W^- b \bar{b} + nj \to \ell^+ \nu \ell^- \bar{\nu} b \bar{b} + nj\,.
\label{eq:bkgttLOnj}
\end{equation}
respectively, where $n$ = 1 or 2 for $t\bar t$ and $W^+W^-$ while $n=1,2,3$ or 4 for $W^\pm c$.  
We reweight the total event rate in each case to the NLO value.   At the level 
of event generation, we apply minimal threshold cuts.  In the CMS case, we reject events with jets
that fall in the range $p_T^j < 10$ GeV and $|\eta^j| < 2.5$, and in the ATLAS case, we reject events with jets that have $p_T^j < 15$ GeV and $|\eta^j| < 4.8$. The cut on $p_T^j$  serves to eliminate collinear 
divergences.

We model detector resolution effects by smearing the final state energies according to:
\begin{equation} 
\frac{\delta E}{E} = \frac{a}{\sqrt{E/{\rm GeV}}} \oplus b
\label{eq:Esmear}
\end{equation}
where we take $a$ = 10\% (50\%) and $b$ = 0.7\% (3\%) for leptons (jets).

\begin{table}[t]
\caption{Kinematic and physics cuts utilized by the ATLAS~\cite{Aad:2009wy} and CMS~\cite{CMS:PPR}
collaborations for the Higgs boson search in the process $gg\to H\to W(\to\ell\nu)W(\to\ell\nu)$.
The SM backgrounds (BKGDs) considered are shown in the second row. Preselection
cuts on the two charged leptons and further physics cuts are presented
in the third and fourth rows, respectively.  Here, $m_{\ell\ell}$ denotes
the invariant mass of the two charged leptons, $\phi_{\ell\ell}$ is 
the angle in the transverse plane between the two charged leptons,
$\met$ is the missing transverse momentum, $p_{T}$ is the transverse momentum, and
$\eta$ is the rapidity.
The cluster transverse mass of the ($\ell\ell,$ $\met$) system is defined as 
$M_{T}^{C}\equiv\sqrt{p_{T}^{2}(\ell\ell)+m^{2}(\ell\ell)}+\met$.
\label{tab: atlas-cms-cut} }

\centering{}\begin{tabular}{c|c|c}
\hline 
 & ATLAS~\cite{Aad:2009wy}  & CMS~\cite{CMS:PPR} \tabularnewline
\hline
\hline 
 & $t\bar t$ & $qq/gg\to WW\to ll$ \tabularnewline
 & $WW$   & $t\bar{t}\to WWbb\to ll$ \tabularnewline
BKGDs  & $W+$jets  & $tWb\to WWb(b)\to ll$ \tabularnewline
 & Single $t$ & $ZW\to lll$ \tabularnewline
 & $Z\to ll$ & $ZZ\to ll$ \tabularnewline
\hline
\hline 
 & Two leptons with opposite charge  & Two tagged leptons with opposite charge \tabularnewline
Preselection  & with $p_{T}>15$ GeV and $|\eta|<2.5$  & with $p_{T}>20$ GeV and $|\eta|<2$ \tabularnewline
 & (crack region $1.37<|\eta|<1.52$)  & \tabularnewline
\hline
\hline 
 &  $12\,{\rm GeV}<m_{ll}<300\,{\rm GeV}$ & $\met>50$ GeV \tabularnewline
 &  $\met>30$ GeV & $\Delta\phi_{ll}<\pi/4$ \tabularnewline
 &  $Z\to\tau\tau$ veto with $|m_{\tau\tau}-m_{Z}|<25\,{\rm GeV}$  & $12\,{\rm GeV}<m_{ll}<40\,{\rm GeV}$ \tabularnewline
Physics &  No jets with $p_{T}>20\,{\rm GeV}$ and $|\eta|<4.8$,  & No jets with $p_{T}>15\,{\rm GeV}$ and $|\eta|<2.5$ \tabularnewline
Cuts  & No b-jets with $p_{T}>15\,{\rm GeV}$  & $30\,{\rm GeV}<{\rm max}\{p_{T}^{l1},p_{T}^{l2}\}<55\,{\rm GeV}$ \tabularnewline
 &  $\Delta\phi_{ll}<\pi/2$ and $M_{T}^{C}<600\,{\rm GeV}$ & $25\,{\rm GeV}<{\rm min}\{p_{T}^{l1},p_{T}^{l2}\}$ \tabularnewline
\hline
\end{tabular}
\end{table}

\section{Comparisons with ATLAS and CMS PPR results at $\sqrt s = 14$~TeV\label{sec:Comparison} }

In this section we compare our expectations with those presented by ATLAS and 
CMS at 14 TeV, focusing on the experimental cuts and their
acceptances.   All of the cross sections in this section
include the decay branching fraction of the Higgs boson into a pair of $W$ bosons and their 
subsequent leptonic decays.  They are summed over two flavors of leptons ($e$ and $\mu$) 
from $W \rightarrow l \nu$, unless specified otherwise.  

\begin{figure}
\includegraphics[scale=0.6]{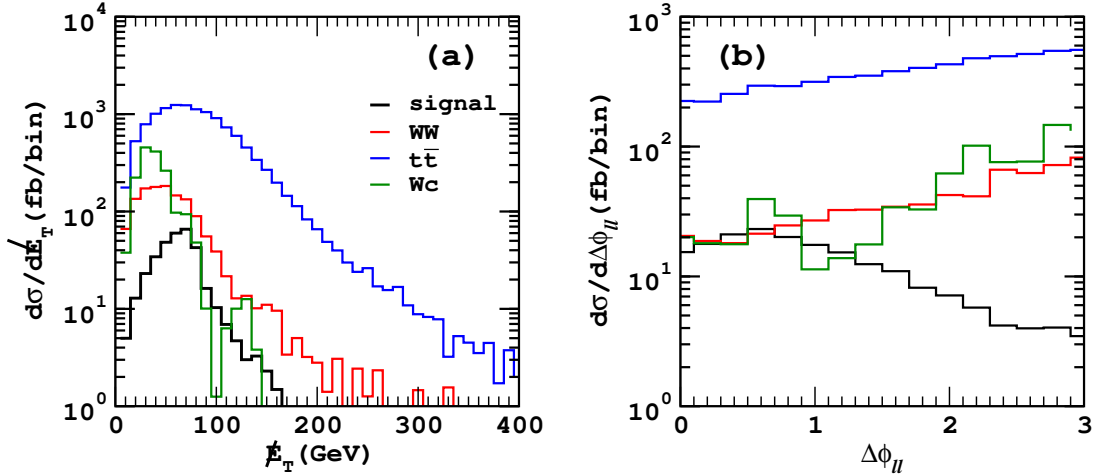}
\caption{Kinematic distributions of $\met$ and $\Delta \phi_{\ell\ell}$
for a 160 GeV Higgs boson with $\sqrt{s} = 14$ TeV, where the black (red, blue,green) curves denote 
the signal ($WW$, $t\bar{t}$, $Wc$) contributions.  Here, ATLAS-motivated 
lepton preselection cuts on $p_T$ and $\eta$ as well as the $m_{\ell\ell}$ cut 
are imposed.  
\label{fig:distro}}
\end{figure}

Somewhat different preselection and physics cuts are chosen by ATLAS~\cite{Aad:2009wy} 
and CMS~\cite{CMS:PPR} in order to suppress the standard model (SM) backgrounds.  These are summarized here in Table~\ref{tab: atlas-cms-cut}.
The cuts are motivated by the different kinematic distributions of the signal and
the backgrounds.  The cut on missing energy rejects  backgrounds like Drell-Yan production 
of a $Z$ boson with $Z\to\ell^{+}\ell^{-}$, which have little or no intrinsic $\met$.   Both the 
signal and  the $t\bar{t}$ and $WW$ backgrounds exhibit
large missing transverse energy associated with the momentum carried off by 
neutrinos.  We display our calculations of the signal and background distributions in 
Fig.~\ref{fig:distro}(a).   The lower cut on the invariant mass of two charged leptons 
removes background from charmed and bottomed mesons, like $J/\psi$
and $\Upsilon$.  The azimuthal angle cut $\Delta\phi_{\ell\ell}$ is valuable for a 
relatively light Higgs boson where the spin correlation of the two $W$-bosons plays an 
important role~\cite{Dittmar:1996ss}, as shown in Fig.~\ref{fig:distro}(b).  The backgrounds favor a large $\Delta \phi_{\ell \ell}$ whereas the signal populates small $\Delta \phi_{\ell \ell}$.  As the Higgs boson mass increases, the separation between signal and background in $\Delta \phi_{\ell \ell}$ is not as strong.

\subsection{ATLAS comparison}

The ATLAS PPR presents the prospects for Higgs boson searches in the 
$g g \rightarrow H X$ channel with subsequent Higgs boson decay 
$H\to WW\to e\mu\nu\nu$.  They include both 
$e^{+}\mu^{-}$ and $e^{-}\mu^{+}$~\cite{Aad:2009wy}, but not the same 
flavor combinations, $\mu^{+}\mu^{-}$ and $e^{+}e^{-}$.    
In this subsection we compare our simulation to the results in the ATLAS report.   
As a prelude, we first attempt to understand all the factors that enter in the 
event rates quoted by ATLAS.   

ATLAS states that the average efficiency to 
reconstruct an electron candidate in the $gg\to H\to WW$ signal events is 
about $60\%$ (before the $p_T$, $\eta$ and isolation cone size $\Delta R$ 
cuts).  The average efficiency to reconstruct a muon candidate is about $94.4\%$.  
Then kinematic cuts for both electron and muon of $p_{T}>15\,{\rm GeV}$
and $\left|\eta\right|<2.5$, and the calorimeter $\Delta R$ isolation requirement,
yield a further net suppression of 83.3\% and 81.7\% for 
electrons and muons, respectively. Starting from a NLO cross section for 
$p p \rightarrow H \rightarrow W W$ via 
gluon fusion for a Higgs boson mass $m_H = 170$~GeV of $19150$~fb and using 
the efficiencies mentioned above, one can reproduce the entry 
of ``lepton selection + $m_{\ell\ell}$'' for $gg\to H$ in Table 3 of ATLAS report, 
\begin{eqnarray}
 &  & \sigma(gg\to H\to WW\to e\mu) \nonumber  \\
 & = & 19150\,{\rm fb}\times\frac{2}{81}\times(0.6\times0.944)\times(0.833\times0.817)\times0.9=166.13 \,\,\rm{fb}.
\label{eq:atlaseq} 
\end{eqnarray}
The factor $0.9$ represents the effect of the cut on the dilepton invariant mass.  

In obtaining our final numbers, we adopt the average efficiencies 
of $60\%$ and $94.4\%$ that ATLAS supplies for electron and muon reconstruction.  
We have no way to compute these.   However, we calculate the effects of all other 
preselection and physics cuts.   Table~\ref{tab: atlas} displays the cut acceptances from 
our simulation along with the ATLAS results.  We show results for the Higgs boson signal 
and the backgrounds.  The cuts are defined in 
Table \ref{tab: atlas-cms-cut}.   After all cuts, we obtain very good agreement with the 
ATLAS study for both signal and backgrounds at 14 TeV for $m_H = 170$~GeV.   

\begin{table}
\caption{Cut acceptance for $m_{H}=170\,{\rm GeV}$ for Higgs boson production via gluon
fusion, with $H\to WW\to e\nu\mu\nu$, at 14~TeV.  The kinematic cuts listed in each row
are applied sequentially. \label{tab: atlas}}
\begin{tabular}{c|cc|cc|cc|cc}
\hline
& \multicolumn{2}{c|}{$H+(0,1,2)j$} & \multicolumn{2}{c|}{$t\bar{t}+(0,1,2)j$} &
\multicolumn{2}{c|}{$WW+(0,1,2)j$} & \multicolumn{2}{c}{$Wc+(0-4)j$}\tabularnewline
\cline{2-9}
& Our  & ATLAS  & Our  & ATLAS  & Our  & ATLAS & Our  & ATLAS \tabularnewline
\hline
\hline
i.d. + $m_{\ell\ell}$  & 100\%  & 100\% & 100\%  & 100\%  & 100\%  & 100\% & 100\%  & 100\% \tabularnewline
\hline
$\met$            & 89\%   & 89\%  & 88\%   & 86\%   & 71\%   & 70\%  & 57\%   & 87\% \tabularnewline
\hline
$Z\to\tau\tau$    & 89\%   & 88\%  & 88\%   & 80\%   & 71\%   & 68\% & 57\%   & 72\% \tabularnewline
\hline
Jet veto          & 37\%   & 37\%  & 0.31\% & 0.23\% & 31\%   & 33\% & 28\%   & 36\% \tabularnewline
\hline
$b$ veto          & 37\%   & 37\%  & 0.31\% & 0.11\% & 31\%   & 33\% & 28\%   & 36\% \tabularnewline
\hline
$\Delta\phi_{\ell\ell}$ and $M_{T}^{C}$
                 & 30\%   & 30\%  & 0.07\% & $(0.04\pm0.03)\%$ &  12\%& $(12\pm0.4)\%$ & 8\% & $(18\pm18)\%$ \tabularnewline
\hline
\end{tabular}
\end{table}

Based on our calculations of the signal and background rates, and applying the 
ATLAS motivated cuts described above, we compute a signal significance $S/\sqrt B = 4.9$ for 
$m_H = 170$ GeV.  This value is in good agreement with the number $4.5$ quoted by ATLAS \cite{Aad:2009wy}.  We understand this difference in terms of our somewhat smaller $W+$ jet background estimate.

\subsection{CMS comparison}

The CMS PPR presents the prospects for Higgs boson searches in the gluon fusion 
channel with subsequent Higgs boson decay $H\to WW\to\ell\ell\nu\nu$
where $\ell=e,\mu$~\cite{CMS:PPR}.  The leptons, whether electrons or muons, are 
required to have $p_{T}>20\,{\rm GeV}$ and $\left|\eta\right|<2$.   The CMS PPR does 
not show cut acceptances for individual cuts, limiting our ability to make as detailed a 
comparison as we do for ATLAS.  We find good agreement between our simulation and 
the CMS results after all cuts are imposed, except in the case of the $t \bar{t}$ background 
where our value is about a factor of $3.5$ below that of CMS. 
This difference may arise from the lower jet 
transverse-momentum threshold taken by CMS.  In the CMS case, we generate events 
with a cut at $10$~GeV at the generator level, and then impose the $15$~GeV cut at the 
event analysis level.  These low values of the cuts may accentuate differences between 
our method and a full parton-showering.  More is written on this discrepancy in the next subsection.  

We do not include the loop-induced background $gg \to WW$ which was considered by CMS.  
At the level of $\sigma_{total}$, this channel makes a relatively small contribution to the 
continuum $WW$ background (less than 5\%).  However, as shown in the CMS PPR, $gg \to WW$ contributes nearly 25\% of the $WW$ rate after all cuts are applied.  The reason for the enhanced contribution after cuts is that the $gg$ subprocess produces a different configuration in  phase space 
from the $q\bar q$ subprocess.  For instance, in the valence-sea $q\bar q$ subprocess, the PDF dependence tends to boost the $WW$ pair to slightly higher rapidity, in contrast with the $gg$ initial state which tends to produce the $WW$ (via the loop induced background or the Higgs boson decay) 
in the central rapidity region.  Therefore, cuts that select on the signal events weaken the rejection of 
the $gg\to WW$ subprocess. 
  
%
\begin{table}
\caption{Comparison to the CMS study for $m_{H}=170\,{\rm GeV}$ at 14~TeV.  
\label{tab:cmsaccept}}
\begin{tabular}{c|cc|cc|cc}
\hline
& \multicolumn{2}{c|}{$H+(0,1,2)j$} & \multicolumn{2}{c|}{$t\bar{t}+(0,1,2)j$} &
\multicolumn{2}{c}{$WW+(0,1,2)j$}\tabularnewline
\cline{2-7}
& Our  & CMS  & Our   & CMS  & Our  & CMS \tabularnewline
\hline
lepton selection  & 100\% & 100\% & 100\% & 100\% & 100\% & 100\%\tabularnewline
\hline
All cuts  & 9.6\%  & 8.8\%  & 0.016\%  & 0.062\%  & 1.16\%  & 1.07\% \tabularnewline
\hline
\end{tabular}
\end{table}

A study of possible $W+$ jets backgrounds is not included in the CMS study reported in their PPR.  
The CMS lepton threshold cut, $p_T^{\ell}>20$ GeV, is harder than ATLAS and is more efficient at removing soft leptons from heavy flavor quark decays in the Higgs boson mass range of interest here~\cite{Sullivan:2006hb,Sullivan:2008ki}.  Nevertheless, we find that the $W + c$ channel can provide a background comparable to (or even larger) than $t\bar{t}$.  Our estimate of its contribution is included in 
our predictions of the Higgs boson search sensitivity at 7 TeV and 10 TeV, reported in 
Section~\ref{sect:sensitivity}.

\subsection{From CMS to ATLAS}

The disagreement of our estimated efficiency for the $t \bar{t}$ background with the CMS value may be contrasted with the good agreement we achieve in the ATLAS case.    As shown in Table~\ref{tab: atlas-cms-cut}, ATLAS imposes slightly different cuts than CMS.   For example, ATLAS requires a harder cut of  $p_{T}>20\,{\rm GeV}$ to veto additional jets while the CMS chooses a softer cut $p_{T}>15\,{\rm GeV}$.    To try to gain some insight into the effects of different cuts, we systematically change the CMS cuts to ATLAS cuts and smoothly transition between the two in Table~\ref{tab:cms2atlas}.  Our approach is to replace a cut imposed by CMS with an equivalent (or nearly equivalent) cut imposed by ATLAS.  In the Table, the cuts are defined as follows:
\begin{itemize}
\item Cut 1 imposes the CMS lepton selection criteria and is the yardstick by which the other acceptances are measured;
\item Cut 2 includes all cuts imposed by CMS and is to be directly compared with Table~\ref{tab:cmsaccept}, but for $M_h=160$ GeV;
\item Cut 3 relaxes the $\met$ cut from $\met>50\,{\rm GeV}\Longrightarrow\met>30\,{\rm GeV}$;
\item Cut 4 relaxes the invariant mass cut $12\,{\rm GeV}<m_{\ell\ell}<40\,{\rm GeV}\Longrightarrow12\,{\rm GeV}<m_{\ell\ell}<300\,{\rm GeV}$;
\item Cut 5 relaxes azimuthal open angle $\Delta\phi_{\ell\ell}<\pi/4\Longrightarrow \Delta\phi_{\ell\ell}<\pi/2$;
\item Cut 6 restores the control region of ATLAS study $M_{T}^{C}<600\,{\rm GeV}$ and removes the 
requirements CMS imposes on the maximum and minimum values of the transverse momenta of 
the leptons;
\item Cut 7 removes the $Z\to\tau\tau$ background, $\left|m_{\tau\tau}-m_{Z}\right|<25\,{\rm GeV}$.  
\end{itemize}
The threshold cuts applied at the analysis level are $p_T^j > 15$ GeV and $|\eta_j| < 2.5$, which are those applied in the CMS study.  The cuts with the largest change in acceptance in going from CMS to ATLAS include Cuts 4, 5, and 6.  Cut 4 was designed to eliminate Drell-Yan background in CMS where all combinations of $e$ and $\mu$ are accepted.  As ATLAS limits their analysis to the opposite flavor $e^\pm \mu^\mp$ channel, the Drell-Yan background is not much of a concern.  In Cut 5, relaxing the 
cut on the opening angle, $\Delta\phi_{\ell \ell}$, increases acceptance.  Such a strict opening angle cut is optimized more for Higgs boson searches through the $WW$ channel near threshold where the final-state leptons are highly correlated.  Of the cuts enumerated above, Cut 6 includes the largest change in cuts and is one of the last steps in going from the CMS cuts to the ATLAS cuts.   It is not a surprise that there is a large shift in acceptance upon changing these cuts.  
In Table~\ref{tab:cms2atlas} we show acceptances for two different generator-level cuts on the jet 
$p_T$ thresholds.  We observe that the acceptances are insensitive to this change, except for the 
$t \bar{t}$ case.

\begin{table}
\caption{Change of the cut acceptances when we switch from the CMS cuts to the
ATLAS cuts.  We apply threshold cuts at the analysis level consistent with the CMS study.  
See the text for cut definitions. We show the acceptances for two different generator-level 
jet $p_T$ thresholds of 10 and 15 GeV.  For the signal process, $M_h = 160$ GeV.
\label{tab:cms2atlas}}
\begin{tabular}{c||c|c|c|c|c|c|c}
\hline
process & Cut 1 & Cut 2 & Cut 3 & Cut4 & Cut5 & Cut6 & Cut7\tabularnewline
\hline
\hline
$gg\to H\to WW$ ($p_{T}^{j}>10\,{\rm GeV}$) & 100\% & 11.2\% & 11.8\% & 13.7\% & 24.7\% & 33.5\% &
33.5\%\tabularnewline
$gg\to H\to WW$ ($p_{T}^{j}\to15\,{\rm GeV}$) & 100\% & 10.7\% & 11.1\% & 12.9\% & 23.2\% & 31.6\% &
31.6\%\tabularnewline
\hline
$t\bar{t}$ ($p_{T}^{j}>10\,{\rm GeV}$) & 100\% & 0.016\% & 0.018\% & 0.028\% & 0.068\% & 0.190\% &
0.184\%\tabularnewline
$t\bar{t}$ ($p_{T}^{j}\to15\,{\rm GeV}$) & 100\% & 0.009\% & 0.010\% & 0.020\% & 0.066\% & 0.164\% &
0.156\%\tabularnewline
\hline
\hline
$WW$ ($p_{T}^{j}>10\,{\rm GeV}$) & 100\% & 1.16\% & 1.18\% & 2.02\% & 5.38\% & 10.48\% &
10.48\%\tabularnewline
$WW$($p_{T}^{j}\to15\,{\rm GeV}$) & 100\% & 1.18\% & 1.19\% & 2.01\% & 5.19\% & 10.08\% &
10.08\%\tabularnewline
\hline
\end{tabular}
\end{table}

After Cut 7 in Table~\ref{tab:cms2atlas}, we expect to obtain results close to those found in our ATLAS analysis, Table \ref{tab: atlas}.  In fact, we see close agreement for the Higgs boson signal and for the $WW$ continuum background.  This agreement is not exact due to differences in the threshold cuts at the generator level and at the analysis level which are initially tailored for CMS.   
Our result for the $t \bar{t}$ background in the Cut 7 column is a factor of two higher than shown in 
Table \ref{tab: atlas}.   We note in this connection that the physics cut for CMS requires no jets with 
$| \eta |< 2.5$, whereas the quiet region for ATLAS  is defined by no jets with $| \eta |< 4.8$.    CMS accepts events which have jets in the region $2.5 < |\eta_j| < 4.8$ that would be vetoed by the ATLAS cuts.  The factor of two is therefore understandable.   Overall, we can explain how our acceptances for CMS transform into our acceptances for ATLAS.  Our results are internally consistent, but we do not have an explanation for the fact that our signal and background acceptances differ from those computed by CMS only in the $t \bar{t} X$ case but compare well with all those computed by ATLAS.  We use both our value and the CMS value of the $t \bar{t} X$ background to bracket our predictions of the Higgs boson search sensitivity at 7 TeV and 10 TeV in Section~\ref{sect:sensitivity}.  Since this background is not dominant, the differences are not large.  

\section{Higgs boson sensitivity at 7 TeV and 10 TeV}
\label{sect:sensitivity}

\subsection{ATLAS}

Having established agreement of the acceptances we compute at 14~TeV with those obtained 
by ATLAS, we extend our analysis to lower LHC cm energies of 7~TeV and 10~TeV.  We use the  
ATLAS cuts shown in Table~\ref{tab: atlas-cms-cut}, assuming the same cuts will apply at lower energies.   Our expectations are presented in Table~\ref{tab:nslhc}.
Here, $\sigma_{tot}$ denotes 
the NLO total cross section of the signal and backgrounds
with the two intermediate $W$-bosons decaying into lepton pairs.  We include
three lepton flavors ($e$, $\mu$, $\tau$) at this stage.  After taking into account
the detector related factors as well as the lepton selection and isolation cuts, 
we obtain the cross section $\sigma_{id}$, the third column in Table~\ref{tab:nslhc}.
In accord with the ATLAS 14~TeV study, 
we only consider the $e\mu$ final state, i.e. both $e^+\mu^-$ and 
$e^- \mu^+$, leading to a decay branching factor of $2/9$.  The optimal ATLAS cuts 
in Table~\ref{tab: atlas-cms-cut} are imposed to
further suppress the SM backgrounds, resulting in the cross section $\sigma_{cut}$
(the fourth column in Table~\ref{tab:nslhc}). 

\begin{table}
\caption{Discovery potential at the Large Hadron Collider with the ATLAS
cuts imposed. Here, $\sigma_{tot}$ denotes the total cross section
(pb) of the signal ($gg\to H\to WW$) and backgrounds, including the $W$-boson decay 
branching ratio into the three flavors of leptons. $\sigma_{id}$ is the cross section
after the lepton reconstruction and isolation, $\sigma_{cut}$ presents
the cross section after all ATLAS cuts are imposed. $\mathcal{A}_{cut}$
is the cut acceptance, defined as $\sigma_{cut}/\sigma_{id}$. \label{tab:nslhc}}

\begin{tabular}{c|c c c c c c}
\hline 
\hline
14 TeV & $\sigma_{tot}$(pb) & $\sigma_{id}$(fb) & $\sigma_{cut}$(fb) & 
$\mathcal{A}_{\rm cut}$ & $\sigma_{S}/\sigma_{B}$ & $\sigma_{S}/\sqrt{\sigma_{B}}$  \tabularnewline   
\hline
$m_{H}=140\,{\rm GeV}$ & 1.50 & 125 & 30.6 & 24.5\% & 18.2\% & 2.36 \tabularnewline
$m_{H}=160\,{\rm GeV}$ & 2.19 & 208 & 61.4 & 29.5\% & 36.7\% & 4.75 \tabularnewline
$m_{H}=180\,{\rm GeV}$ & 1.85 & 183 & 47.9 & 26.2\% & 28.6\% & 3.70 \tabularnewline
$m_{H}=200\,{\rm GeV}$ & 1.21 & 110 & 21.8 & 19.8\% & 13.0\% & 1.68 \tabularnewline
$t\bar{t}+(0,1,2)j$    &89.7 & 6350 & 4.64 & 0.073\% & - & - \tabularnewline
$WW+(0,1,2)j$          &12.7 &  737 & 87.0 & 11.8\%  & - & - \tabularnewline
$Wc+(0,1,2,3,4)j$      & 2820  &  945 &75.8 & 8.02\%  & - & - \tabularnewline
\hline 
\hline
10 TeV & $\sigma_{tot}$(pb) & $\sigma_{id}$(fb) & $\sigma_{cut}$(fb) & 
$\mathcal{A}_{\rm cut}$ & $\sigma_{S}/\sigma_{B}$ & $\sigma_{S}/\sqrt{\sigma_{B}}$  \tabularnewline   
\hline
$m_{H}=140\,{\rm GeV}$ & 0.773 &  63.8 & 16.7 & 26.2\% & 13.3\% & 1.49 \tabularnewline
$m_{H}=160\,{\rm GeV}$ & 1.12 & 107 & 33.9 & 31.7\% & 26.9\% & 3.02 \tabularnewline
$m_{H}=180\,{\rm GeV}$ & 0.939 & 92.5 & 24.8 & 26.8\% & 19.7\% & 2.21 \tabularnewline
$m_{H}=200\,{\rm GeV}$ & 0.610 &  55.5 & 11.3 & 20.4\% &  8.98\% & 1.00 \tabularnewline
$t\bar{t}+(0,1,2)j$    &41.5 &3070 &  2.46 & 0.080\% & - & - \tabularnewline
$WW+(0,1,2)j$          & 7.99 & 506 & 66.3 & 13.1\%  & - & - \tabularnewline
$Wc+(0,1,2,3,4)j$      & 1670  & 571 & 57.1 & 10.0\%  & - & - \tabularnewline
\hline 
\hline
7 TeV & $\sigma_{tot}$(pb) & $\sigma_{id}$(fb) & $\sigma_{cut}$(fb) & 
$\mathcal{A}_{\rm cut}$ & $\sigma_{S}/\sigma_{B}$ & $\sigma_{S}/\sqrt{\sigma_{B}}$  \tabularnewline   
\hline 
$m_{H}=140\,{\rm GeV}$ & 0.397 &  31.5 &  8.66 & 27.5\% &  9.9\% & 0.923 \tabularnewline
$m_{H}=160\,{\rm GeV}$ & 0.563 &  53.1 & 17.3 & 32.5\% & 19.8\% & 1.85 \tabularnewline
$m_{H}=180\,{\rm GeV}$ & 0.458 &  44.9 & 12.8 & 28.6\% & 14.6\% & 1.37 \tabularnewline
$m_{H}=200\,{\rm GeV}$ & 0.291 &  26.6 &  5.93 & 22.3\% &  6.76\% & 0.633 \tabularnewline
$t\bar{t}+(0,1,2)j$    &17.6 & 1340 & 1.34 & 0.100\% & - & - \tabularnewline
$WW+(0,1,2)j$          & 4.73 &  337 &48.2 & 14.3\%  & - & - \tabularnewline
$Wc+(0,1,2,3,4)j$     &905 &  322 &38.0 & 11.8\%  & - & - \tabularnewline
\hline
\hline

\end{tabular}
\end{table} 

We note that, at 14~TeV, the optimal cuts work best for a Higgs boson
in the mass range of 140 GeV to 180 GeV, yielding cut acceptances  
$\mathcal{A}_{cut}(=\sigma_{cut}/\sigma_{id})$ around 25-30\%.
However, the cut acceptance decreases below 20\% for a heavy Higgs boson,
say $m_H>200~\rm{GeV}$.   This decrease is caused mainly by the cut 
$\Delta\phi_{\ell\ell} < \pi/2 $
as the two charged leptons from the two on-shell $W$-bosons peak near 
$\Delta\phi_{\ell\ell} \sim \pi$ at high mass.

\begin{figure}
\includegraphics[scale=0.4,clip]{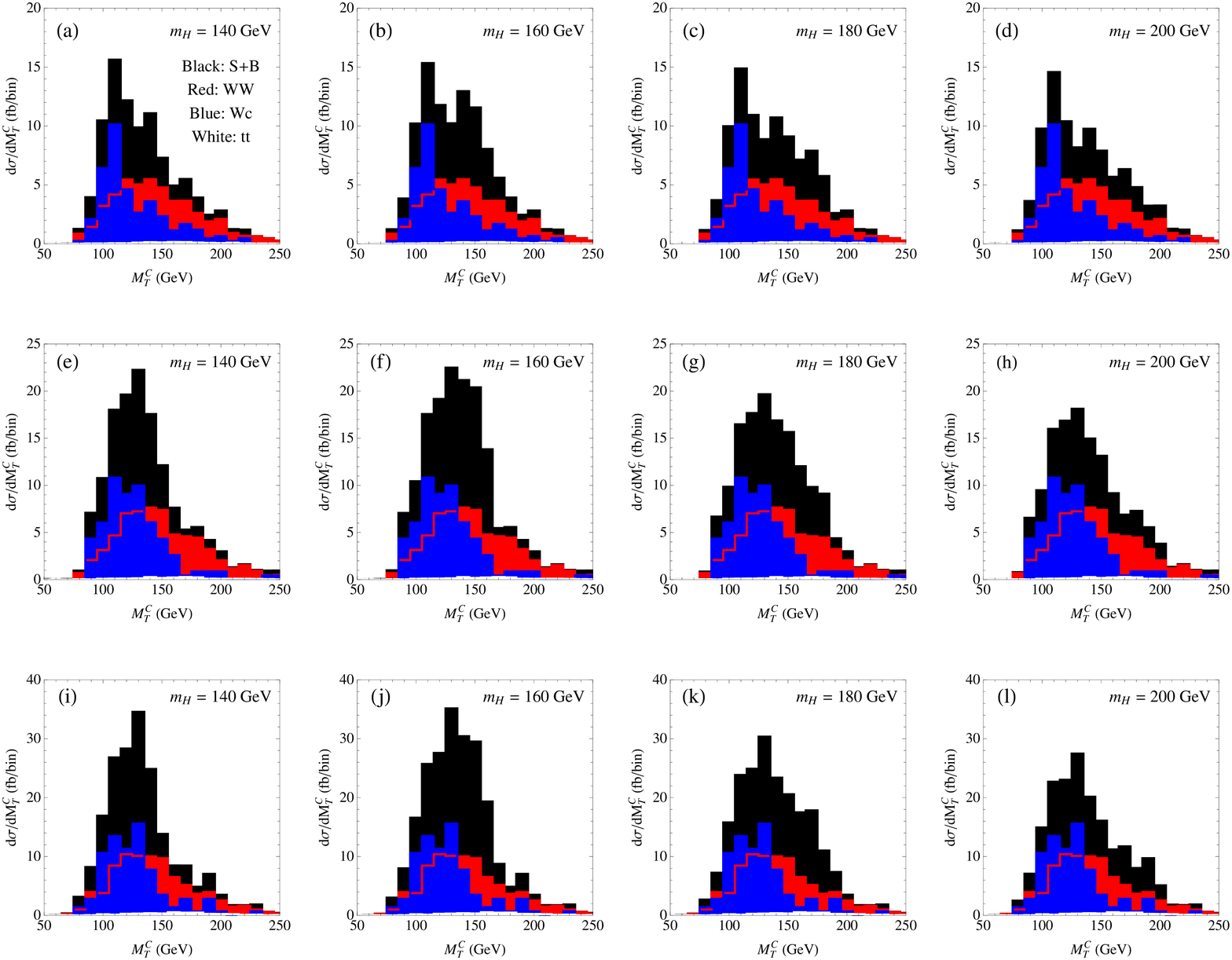}
\caption{Distributions of the cluster transverse mass $M_T^C$
for 7~TeV (a-d), 10~TeV (e-h) and 14~TeV (i-l), 
where the black region denotes 
the signal plus background while the red, blue and white regions
denote the $WW$, $Wc$ and $t\bar{t}$ backgrounds, respectively. 
Here, all ATLAS cuts are imposed.
\label{fig:mt_cluster}}
\end{figure}

After imposing all cuts we find that the signal is below the backgrounds 
where the $WW$ continuum is the leading background and $W$+jets is 
the sub-leading background.  
In Fig.~\ref{fig:mt_cluster}, we show the distribution in $\ell^+\ell^- \met$ 
cluster transverse mass, $$M_{T}^{C}\equiv\sqrt{p_{T}^{2}(\ell\ell)+m^{2}(\ell\ell)}+\met,$$
for the Higgs boson signal and three principal backgrounds
at LHC energies of 7, 10, and 14~TeV.  
The $Wc$ background is most important at low values of $M_T^C$ while 
the $WW$ continuum tends to be important at large values of $M_T^C$.
The top quark pair background is quite small owing to the jet veto cut.

At 14~TeV the signal to background ratio ($S/B\equiv\sigma_{s}/\sigma_{B}$) 
increases from 18\% to 28\% when the Higgs boson mass increases from 140~GeV
to 180~GeV, but it decreases to 13\% for a 200 GeV Higgs boson. 
The decrease is the net effect 
of the cut acceptance and branching ratio $Br(H\to WW)$. The latter drops from 
94\% to 73\% when the Higgs boson mass increases from 180 GeV to 200 GeV, owing to
opening of the $ZZ$ mode.
We present the significance in the last column of Table~\ref{tab:nslhc}
assuming an integrated luminosity ($\mathcal{L}$) of $1~{\rm fb}^{-1}$.   

Upon lowering the cm energy we find that the cut acceptances increase for both
the signal and backgrounds.
The signal cross section decreases more than the $WW$
background cross section.  
As a result, the $S/B$ ratios at 10 TeV
and 7 TeV are less than those at 14 TeV. The lower cm energy, the smaller
$S/B$ ratio and the signal significance. The latter is defined as
\begin{equation}
\frac{S}{\sqrt{B}} = \frac{\sigma_S}{\sqrt{\sigma_B}}\times \sqrt{\mathcal{L}}. 
\end{equation}

\begin{table}
\caption{Similar to Table~\ref{tab:nslhc},  discovery potential at the Large Hadron Collider with CMS
cuts imposed for 1 ${\rm fb}^{-1}$.  We provide two sets of values for the $t \bar{t} X$ background, our 
computed value and in parentheses the value attributed to CMS.  \label{tab:nslhc-cms}}

\begin{tabular}{c|c c c c c c}
\hline 
\hline
14 TeV & $\sigma_{tot}$(pb) & $\sigma_{id}$(fb) & $\sigma_{cut}$(fb) & 
$\mathcal{A}_{\rm cut}$ & $\sigma_{S}/\sigma_{B}$ & $\sigma_{S}/\sqrt{\sigma_{B}}$  \tabularnewline   
\hline
$m_{H}=140\,{\rm GeV}$ & 1.50 & 160 & 7.15 & 4.47\% & 36.0\% (27.9\%) & 1.60 (1.41) \tabularnewline
$m_{H}=160\,{\rm GeV}$ & 2.19 & 340 & 37.7 & 11.1\% & 190\% (147\%) & 8.46 (7.45) \tabularnewline
$m_{H}=180\,{\rm GeV}$ & 1.85 & 309 & 17.8 & 5.75\% & 89.6\% (69.5\%)& 3.99 (3.52) \tabularnewline
$m_{H}=200\,{\rm GeV}$ & 1.21 & 229 & 4.23 & 1.85\% & 21.3\% (16.5\%) & 0.949 (0.836) \tabularnewline
$t\bar{t}+(0,1,2)j$    &89.7 & 13400 & 2.28 (8.04) & 0.017\% (0.06\%)& - & - \tabularnewline
$WW+(0,1,2)j$          &12.7 &  913 &11.3 & 1.24\%  & - & - \tabularnewline
$Wc+(0,1,2,3,4)j$      & 2820  &  3430 & 6.28 & 0.183\%  & - & - \tabularnewline
\hline 
\hline
10 TeV & $\sigma_{tot}$(pb) & $\sigma_{id}$(fb) & $\sigma_{cut}$(fb) & 
$\mathcal{A}_{\rm cut}$ & $\sigma_{S}/\sigma_{B}$ & $\sigma_{S}/\sqrt{\sigma_{B}}$  \tabularnewline   
\hline
$m_{H}=140\,{\rm GeV}$ & 0.773 &  82.0 & 4.23 & 5.17\% & 29.6\% (26.3\%) & 1.12 (1.05)\tabularnewline
$m_{H}=160\,{\rm GeV}$ & 1.12 & 173 & 19.6 & 11.4\% & 137\% (122\%) & 5.19 (4.88) \tabularnewline
$m_{H}=180\,{\rm GeV}$ & 0.939 & 161 & 9.63 & 5.98\% & 67.4\% (59.8\%) & 2.55 (2.40) \tabularnewline
$m_{H}=200\,{\rm GeV}$ & 0.610 &  116 & 2.24 & 1.93\% &  15.7\% (13.9\%) & 0.593 (0.558) \tabularnewline
$t\bar{t}+(0,1,2)j$    &41.5 & 6500 &  0.715 (2.54) & 0.011\% (0.039\%) & - & - \tabularnewline
$WW+(0,1,2)j$          & 7.99 & 659 & 8.17 & 1.24\%  & - & - \tabularnewline
$Wc+(0,1,2,3,4)j$      & 1670  & 1930 & 5.40 & 0.280\%  & - & - \tabularnewline
\hline 
\hline
7 TeV & $\sigma_{tot}$(pb) & $\sigma_{id}$(fb) & $\sigma_{cut}$(fb) & 
$\mathcal{A}_{\rm cut}$ & $\sigma_{S}/\sigma_{B}$ & $\sigma_{S}/\sqrt{\sigma_{B}}$  \tabularnewline   
\hline 
$m_{H}=140\,{\rm GeV}$ & 0.397 &  39.1 &  1.99 & 5.08\% &  23.2\% (21.6\%) & 0.680 (0.655) \tabularnewline
$m_{H}=160\,{\rm GeV}$ & 0.563 &  88.4 & 10.6 & 12.0\% & 124\% (115\%)& 3.62 (3.49) \tabularnewline
$m_{H}=180\,{\rm GeV}$ & 0.458 &  79.3 & 4.82 & 6.08\% & 56.3\% (52.2\%) & 1.65 (1.59) \tabularnewline
$m_{H}=200\,{\rm GeV}$ & 0.291 &  57.1 &  1.17 & 2.05\% &  13.7\% (12.7\%) & 0.400 (0.385) \tabularnewline
$t\bar{t}+(0,1,2)j$    &17.6 & 2890 & 0.260 (0.925) & 0.009\% (0.032\%) & - & - \tabularnewline
$WW+(0,1,2)j$          & 4.72 &  428 & 5.31 & 1.24\%  & - & - \tabularnewline
$Wc+(0,1,2,3,4)j$     &905 &  1050 & 2.99 & 0.285\%  & - & - \tabularnewline
\hline
\hline

\end{tabular}
\end{table} 

\subsection{CMS}

The results of our analysis of CMS expectations at LHC cm energies of 7~TeV and 10~TeV  
are shown in Table~\ref{tab:nslhc-cms}.   We use the CMS 
cuts shown in Table~\ref{tab: atlas-cms-cut}, assuming the same cuts will apply at lower energies.   
Here, $\sigma_{tot}$ denotes the NLO total cross section of the signal and backgrounds, 
including the lepton-pair branching fractions for decay of the two intermediate $W$-bosons
into three lepton flavors ($e$, $\mu$, $\tau$).   The identified cross section $\sigma_{id}$ is obtained 
from  $\sigma_{tot}$ after lepton selection and isolation cuts are imposed on the events we generate, 
as defined in Sec.~\ref{sec:Comparison}.  Recall that CMS averages over all decay modes including 
$ee, \mu\mu$, and $e\mu$.   We also include the CMS trigger efficiencies (``L1 + HLT'') in $\sigma_{id}$.  
Since we have no way of computing these efficiencies, we assume that they are independent of 
the cm energy and use the values in the CMS PPR ({\em c.f.} Fig. 10.12 on page 1276).  
For the signal, the value of the trigger efficiency  depends on the Higgs boson mass and varies from 0.50 to 0.63 for the range of masses considered here.  For the backgrounds, the corresponding efficiencies are 0.67 for $t\bar{t}$ and 0.52 for $WW$.  

The CMS analysis cuts in Table~\ref{tab: atlas-cms-cut} are imposed, resulting in the cross section 
$\sigma_{cut}$ (the fourth column in Table~\ref{tab:nslhc-cms}). 
For the $t\bar{t}$ background, we list two values of $\sigma_{cut}$: one corresponds to the value obtained from our calculation, while the other in parentheses is the larger value quoted in the 
CMS PPR.  As pointed 
out above, the two differ by roughly a factor of 3.5 at a cm energy of 14 TeV.  Since we are unsure of the cause of this discrepancy, we give results for both cut efficiencies as a way of bracketing our uncertainty.  At the lower cm energies, we simply rescale our computed value of $\sigma_{cut}$ for the $t\bar{t}$ background by a factor of 3.5 to obtain what we assume CMS would obtain.  

We include a $W$+jets background contribution (computed as $W+c$) in addition to $t\bar{t}$ 
and $WW$.  This channel can provide a background comparable to (or even larger) than $t\bar{t}$.  Since CMS did not include this background in their analysis, there is no quote for the ``L1+HLT'' trigger efficiency for this channel.  The numbers shown in  Table~\ref{tab:nslhc-cms} for $W$+jets assume perfect efficiency and, thus, are an overestimate of the true rate. 

For the signal, the cut acceptances  $\mathcal{A}_{cut}(=\sigma_{cut}/\sigma_{id})$ for CMS tend to 
be smaller than for ATLAS, in the 5 - 10\% range versus 25-30\% at 14~TeV  However,  the signal to 
background fractions are higher, exceeding 100\% at $m_H = 160$~GeV.   Our calculated signal significance is found in the last column of Table~\ref{tab:nslhc-cms} for an assumed integrated luminosity ($\mathcal{L}$) of $1~{\rm fb}^{-1}$.  For $m_H = 160$~GeV, we find a signal significance 
of 7.4 to 8~$\sigma$ at 14~TeV, depending on how the $t \bar{t}$ background is estimated.  This value 
drops to about 5~$\sigma$ at 10~TeV and to about 3~$\sigma$ at 7~TeV.  The significance drops off on both sides of $m_H = 160$~GeV at all energies.   

We find reasonable agreement of our predicted significance at 14 TeV to that of CMS for $m_H = 160$~GeV.  If  the $gg \to WW$ contribution is omitted, the total signal and background ($t\bar{t} + WW$) cross sections  in the CMS PPR (after all cuts) are $\sigma_S = 42$ fb and $\sigma_B = 21.8$ fb, and the signal significance is then $\sigma_S/\sqrt{\sigma_B} = 9.00$.   If we remove the $W+c$ background from our analysis,  our total cross sections (after all cuts) are $\sigma_S = 35.4$ fb and $\sigma_B = 19.3$ fb, and our signal significance is $\sigma_S/\sqrt{\sigma_B} = 8.05$.  The difference between the two expectations, roughly $\sim$ 10\%, seems to us well within the theoretical uncertainties.   The agreement at 14~TeV lends credibility to our estimates at the lower energies.  

There are uncertainties associated with our values of the cut acceptance $\mathcal{A}_{cut}$
and corresponding signal significance $\sigma_S/\sqrt{\sigma_B}$ in Tables~\ref{tab:nslhc} 
and~\ref{tab:nslhc-cms}.   The most obvious uncertainties can be traced to the choice of PDFs and 
the renormalization and factorization scales.  We also use inclusive NLO K factors, rather than 
K factors that apply in the restricted part of phase space after analysis cuts are applied.  Some of 
these uncertainties are reduced in ratios such as $\mathcal{A}_{cut}$.  The uncertainties presented 
in Figs.~\ref{fig:significanceATLAS-CMS} --~\ref{fig:lumiratio}  are based solely on the statistics of the samples of events that we generate and are no doubt an underestimate of the full 
uncertainty.  

\subsection{ATLAS and CMS Comparison} 

Comparing  Table~\ref{tab:nslhc-cms} and Table~\ref{tab:nslhc}, we see that considerably larger 
signal significance at $m_H = 160$~GeV is obtained with the CMS cuts.  We attribute this difference to the effects of the different analysis cuts in the two cases, particularly the cuts on $m_{ll}$ and 
$\Delta\phi_{ll}$.   The advantage of these cuts diminishes for values of $m_H$ below and above 
$160$~GeV.  

\begin{figure}
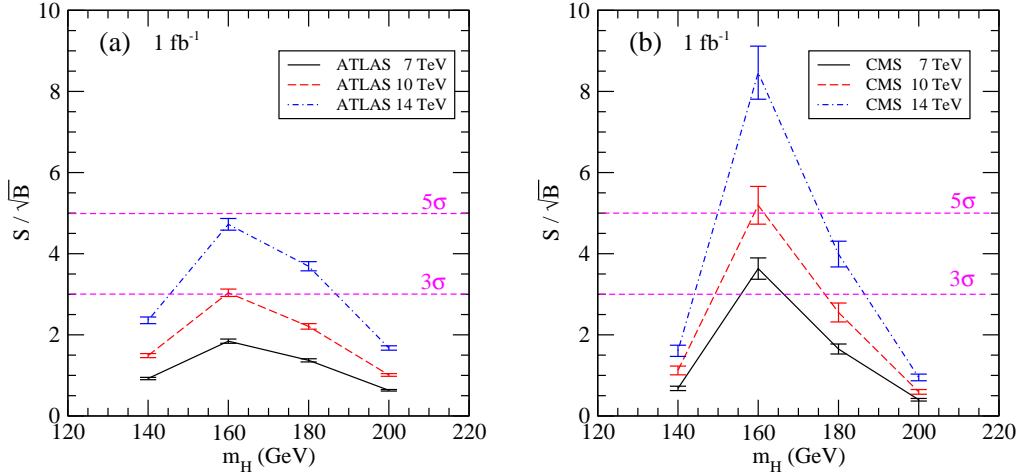

\includegraphics[clip,scale=0.5]{figures/ATLAS_significance}~~~~
\includegraphics[clip,scale=0.5]{figures/CMS_significance}
\caption{Signal significance as a function of $m_{H}$ at ATLAS and CMS with $1\,{\rm fb}^{-1}$
luminosity.} 
\label{fig:significanceATLAS-CMS}
\end{figure}

In Fig.~\ref{fig:significanceATLAS-CMS}, we display our computed signal significance for 
$1~{\rm fb}^{-1}$ of integrated luminosity as a function of Higgs boson mass.  Results are shown at three values of the LHC energy based on the ATLAS and CMS cuts.  Except at $m_H = 160$~GeV, the expectations are similar for the two sets of cuts.  At $m_H = 160$~GeV, we see that ATLAS may achieve nearly $5~\sigma $ significance with $1~{\rm fb}^{-1}$  at 14~TeV, and about $3~\sigma$ and almost 
$2~\sigma$ at $10$~TeV and $7$~TeV, respectively.  The corresponding CMS numbers are 
roughly $8~\sigma$, $5~\sigma$, and $3.5~\sigma$ at 14, 10, and 7~TeV, respectively.  

\begin{figure}
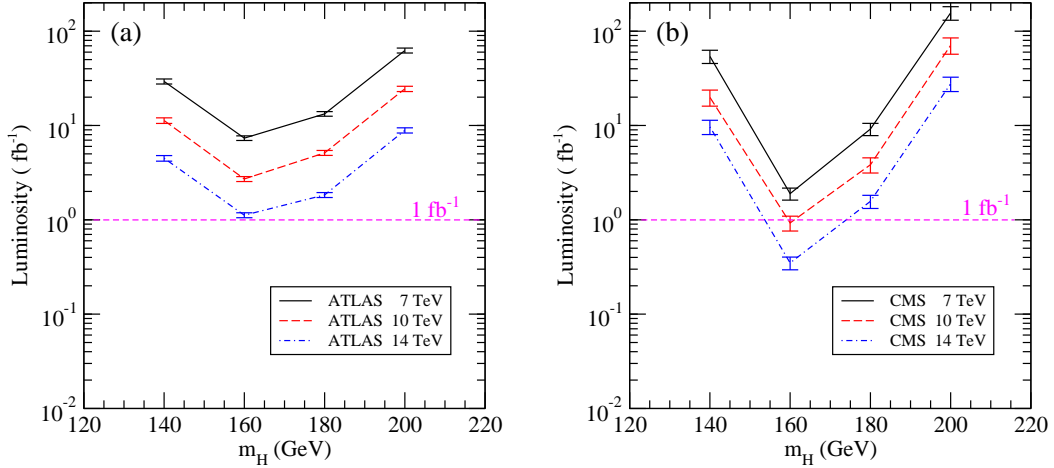

\includegraphics[clip,scale=0.5]{figures/ATLAS_5sigmalumi}~~~~
\includegraphics[clip,scale=0.5]{figures/CMS_5sigmalumi}
\caption{Integrated luminosity required for $5\sigma$ discovery as a function of $m_{H}$ 
in the ATLAS and CMS cases.}
\label{fig:luminosityATLAS-CMS}
\end{figure}

In Fig.~\ref{fig:luminosityATLAS-CMS} we present our calculation of  the integrated luminosity required to achieve  $5\sigma$ discovery as a function of $m_{H}$.  In the ATLAS case, we see that 
$1~{\rm fb}^{-1}$ is essentially sufficient for $m_H = 160$~GeV at 14~TeV, but increases to 
about $3~{\rm fb}^{-1}$ and $8~{\rm fb}^{-1}$ are needed at 10 and 7~TeV, respectively.  
With the CMS analysis cuts, $1~{\rm fb}^{-1}$ is more than sufficient for $m_H = 160$~GeV at 14~TeV, 
and is sufficient at 10~TeV, but an increase to about $2~{\rm fb}^{-1}$ 
is needed at 7~TeV.  

To reproduce the significance at 14 TeV,  a larger luminosity is needed 
at the lower energies to compensate the additional suppression of signal cross section 
compared to the background cross section. 
The enhancement factor of the luminosity is 
\begin{equation}
\frac{\mathcal{L}_{i}}{\mathcal{L}_{14}} = 
\left[
\left(\frac{\sigma_S}{\sqrt{\sigma_B}}\right)_{14} 
~\biggr/
\left(\frac{\sigma_S}{\sqrt{\sigma_B}}\right)_{i}
\right]^2. 
\end{equation}

\begin{figure}
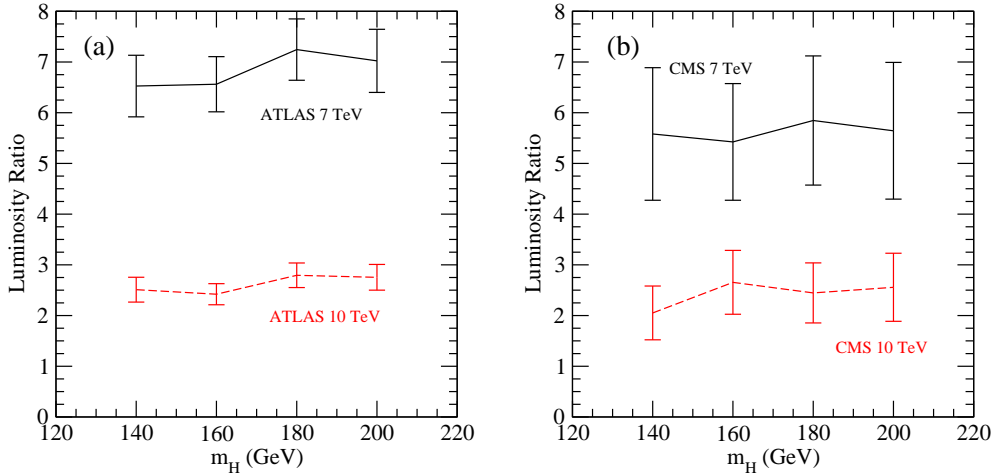

\includegraphics[clip,scale=0.5]{figures/ATLAS_lumiratio_err}~~~~
\includegraphics[clip,scale=0.5]{figures/CMS_lumiratio_err}
\caption{Factors by which the integrated luminosity must be increased at LHC energies of $7$ and $10$~TeV to achieve the same signal significance as at $14$~TeV.
\label{fig:lumiratio}}
\end{figure}

For the two experiments, Fig.~\ref{fig:lumiratio} shows the factors by which the luminosity must be increased at lower energies to discover a Higgs boson with the same significance as at 14 TeV.  Over the range of Higgs boson masses considered, one would need to increase ${\cal{L}}$ by a factor of roughly 2.5 at $10$~TeV for the cuts we associate with ATLAS and CMS.   At $7$~TeV, the factor is 
$\sim 5$ in the CMS case and $\sim 6.5$ in the ATLAS case.   These predictions are relatively 
unaffected by the different values of ${\cal{A}}_{cut}$ for the $t\bar{t}$ background in the CMS case.
The uncertainties are about 9\% at 10~TeV and 8.6\% at 7~TeV in the ATLAS case and close 
to 25\% for both energies in the CMS case.  The more restrictive analysis cuts in the CMS case 
({\em c.f.} our Table~\ref{tab: atlas-cms-cut}) lead to a smaller event sample and therefore larger statistical uncertainties.  

\section{Exclusions -- Tevatron and LHC}
\label{sect:exclude}

\begin{figure}
\includegraphics[clip,scale=0.5]{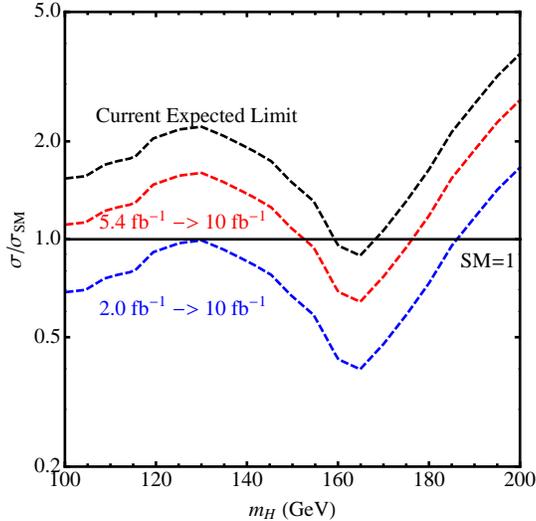}
\caption{Current 95\% exclusion limits for a SM Higgs boson from the combined CDF and D0 study, and projected exclusion limits with $10~{\rm fb}^{-1}$ of integrated luminosity at the Tevatron. } 
\label{fig:tevexcludes}
\end{figure}

The $H \rightarrow W W $ and other decay modes are used in the recent CDF and D0 combined fit 
of Tevatron data to exclude the mass range $163\,{\rm GeV}<m_{H}<166\,{\rm {GeV}}$ at 95\% 
C.~L.~\cite{Phenomena:2009pt}.   The integrated luminosity differs among the various production and decay modes in the Tevatron study,  ranging from 2.0 fb$^{-1}$ to 5.4 fb$^{-1}$. 
In Fig.~\ref{fig:tevexcludes}, we sketch the current combined Tevatron limit on Higgs boson production 
in units of the SM cross section, under the assumption of an exclusion. 
To estimate naively the sensitivity that Tevatron studies will achieve with 10 fb$^{-1}$ by perhaps the 
end of 2011, we multiply the current expected limit by $\sqrt{{\cal L}_{current} /{\cal L}_{projected}}$.  We choose to scale the expected limit rather than the observed limit since it is based on a larger event sample and less subject to statistical fluctuations.  Scaling the observed limit is more sensitive to the fluctuations present in observed events, expected to average out if no signal is present.   We provide 
two possible extrapolations, from either $2.0~{\rm fb}^{-1}$ or $5.4~{\rm fb}^{-1}$ of integrated luminosity.   The Tevatron analyses of the $WW\to \ell^+ \nu \ell^- \bar \nu$ channel include up to 
5.4~fb$^{-1}$ of data.  Therefore, the limit for Higgs boson masses above 160 GeV is expected to follow the projected $5.4~{\rm fb}^{-1} \to 10\text{ fb}^{-1}$ curve, yielding an exclusion of the range 153 to 177~GeV (based on statistics alone).    ÊA detailed analysis of a combined Tevatron exclusion limit with up to $4.2~{\rm fb}^{-1}$ of integrated luminosity~\cite{:2009pt} is provided in 
Ref.~\cite{Draper:2009fh} in which individual channels are scaled by the respective ratios 
$\sqrt{{\cal L}_{current} /{\cal L}_{projected}}$ and then combined.  We observe that the projected 
$10~{\rm fb}^{-1}$ exclusion limit in Ref.~\cite{Draper:2009fh} and our  $5.4~{\rm fb}^{-1} \rightarrow 
 10~{\rm fb}^{-1}$ projection in Fig.~\ref{fig:tevexcludes} are almost identical, supporting the validity 
 of the projections in the relevant mass range $\sim 150$ to $\sim 180$~GeV .  An increase in 
 efficiency can improve the mass exclusion limits~\cite{Draper:2009fh}.  

\begin{figure}
\includegraphics[clip,scale=0.5]{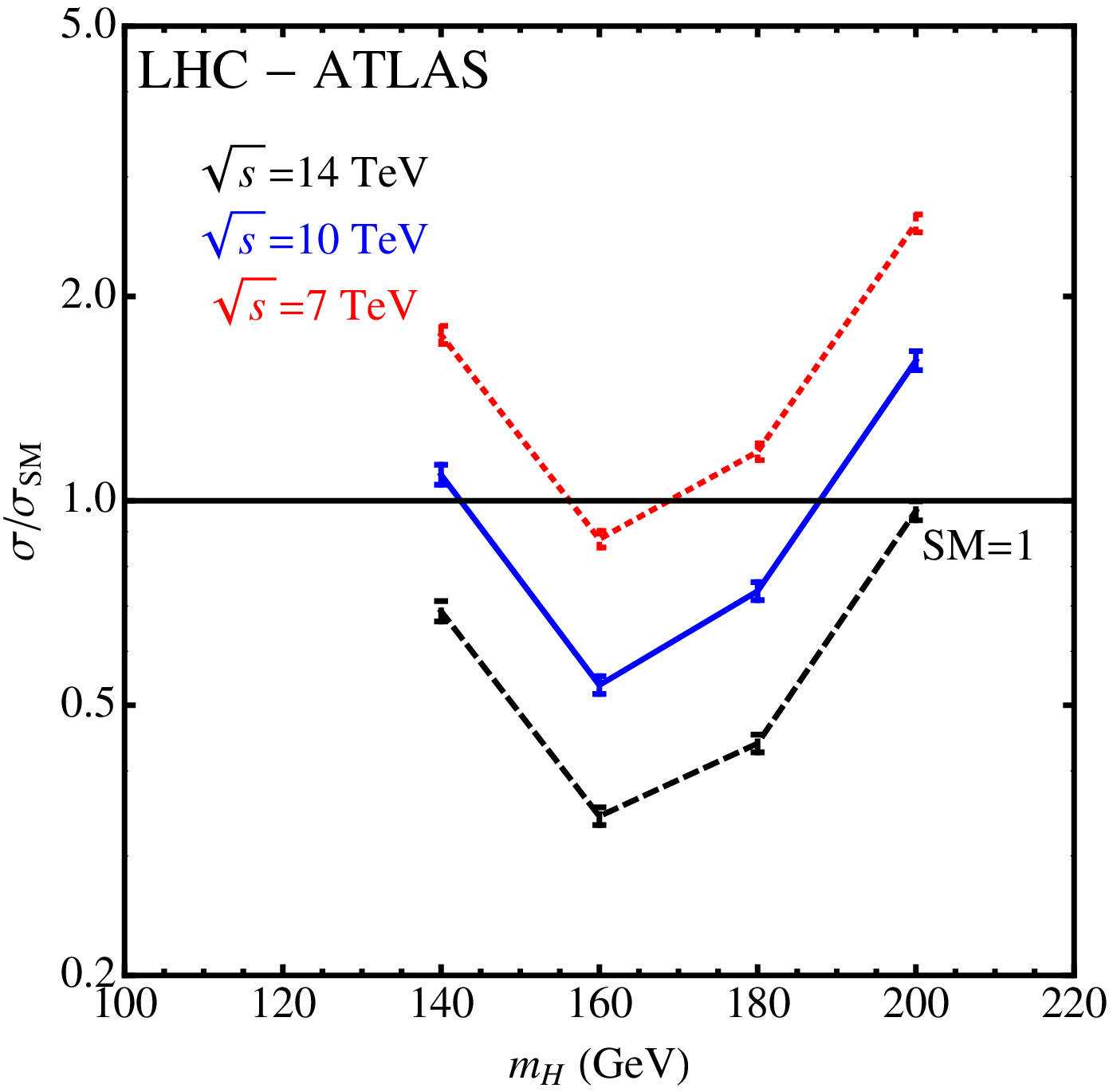}~~~~
\includegraphics[clip,scale=0.5]{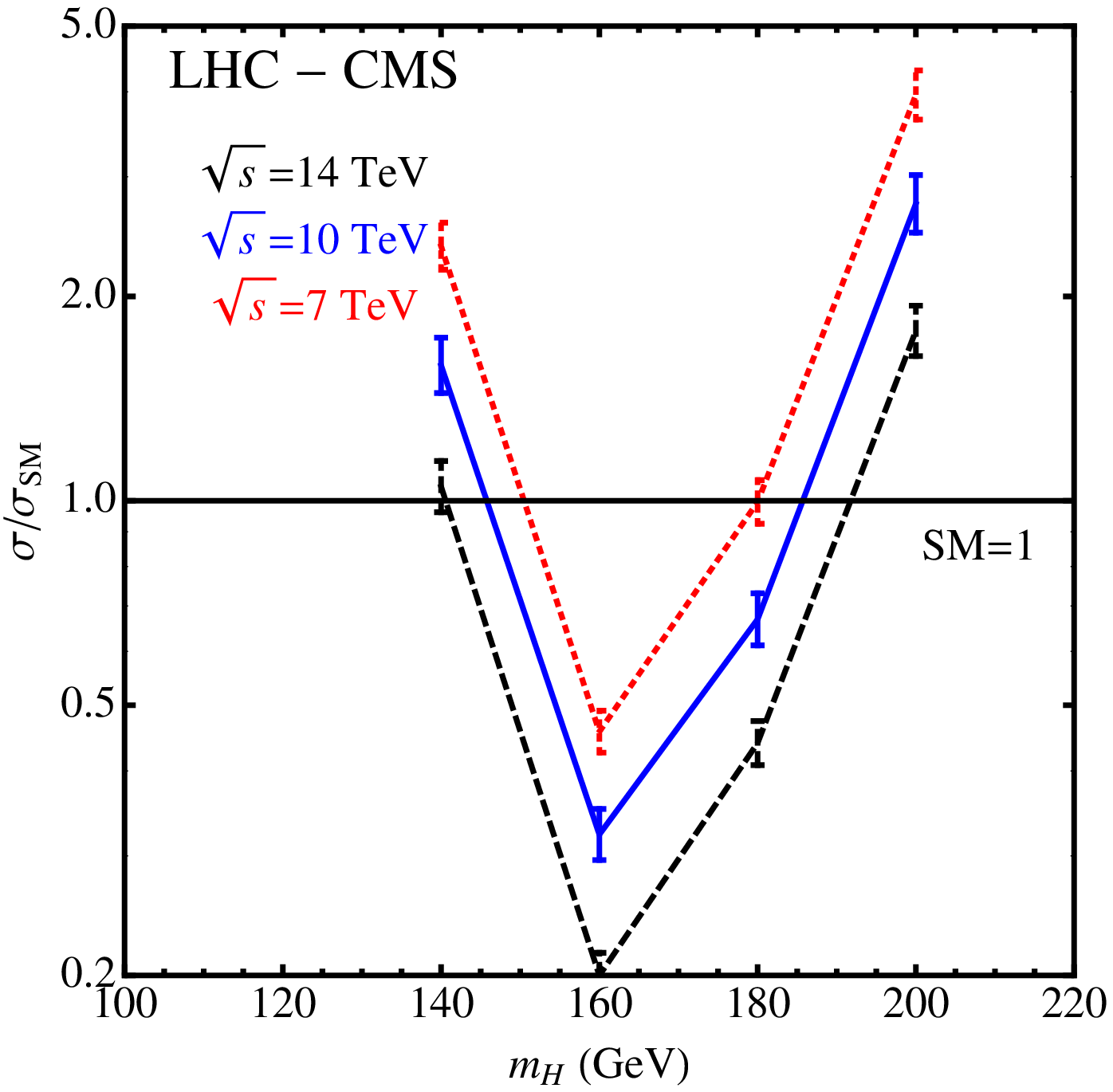}
\caption{Estimated exclusion limits for a SM Higgs boson produced via gluon fusion and decaying via 
$WW \rightarrow l l X$ with $1~{\rm fb}^{-1}$ of integrated luminosity for the ATLAS and CMS experiments for three LHC  energies.} 
\label{fig:LHCexcludes}
\end{figure}

To compare with the expected sensitivity of the Tevatron, our LHC results can be recast as limits on the Higgs boson production cross section assuming no signal is found.  Based on Poisson
statistics, we determine the signal cross section at the LHC that is consistent at 95\% C.L. with the
calculated background for the ATLAS and CMS cuts Refs.~\cite{Aad:2009wy} and~\cite{CMS:PPR}
(using our calculation of $t\bar t$).  These cross sections can be divided by the SM Higgs boson production cross sections to obtain the limits shown in Fig.~\ref{fig:LHCexcludes}.   The results show that with $1~{\rm fb}^{-1}$ of integrated luminosity at 7~TeV, CMS may be able to exclude $m_H$ values of 160 and 180~GeV (and perhaps points in between), while ATLAS may be able to exclude 160~GeV.  The difference between the two experiments is traced to the different set of analysis cuts, described 
in Refs.~\cite{Aad:2009wy} and~\cite{CMS:PPR} and summarized in our Table~\ref{tab: atlas-cms-cut}. Figure~\ref{fig:LHCexcludes} suggests that the exclusion range at the LHC may extend below 
160~GeV.  However, since our calculations are done at fixed values of the Higgs boson mass, we refrain from making statements for values other than those at which we have explicit results.   
The figures show that with $1~{\rm fb}^{-1}$ of 
integrated luminosity  at 7~TeV the ATLAS exclusion range should be comparable to the current Tevatron exclusion range,  and that the CMS exclusion range should be comparable to a Tevatron 
study with  $10~{\rm fb}^{-1}$.   Comparison of Figs.~\ref{fig:tevexcludes} and~\ref{fig:LHCexcludes} 
indicates that the Tevatron experiments should remain competitive through 2011 and perhaps beyond provided they achieve analyses based on $10~{\rm fb}^{-1}$ of integrated luminosity.  

\section{Conclusions}
\label{sect:conclusions}

In this paper, we provide Higgs boson discovery prospects for early LHC running at energies of 7 and 
10~TeV for the $gg\to H\to W^+ W^- \to \ell^+\ell^- + \met$ channel.  Our estimates of the Higgs boson signal and backgrounds are obtained from parton-level simulations. We apply the same cuts used by the CMS~\cite{CMS:PPR} and ATLAS~\cite{Aad:2009wy} collaborations, and we verify that our acceptances for the signal, $W^+W^-$  and $W^\pm + n$ jets channels agree with those of CMS and ATLAS.  Our acceptance for the $t\bar t$ process matches well when compared to the ATLAS results, but it is in less good agreement with the CMS results.  This difference may be understood in terms of the low jet $p_T$ threshold taken by CMS where a full parton shower simulation may be necessary.   We obtain good agreement with the ATLAS and CMS collaboration estimates of signal significance at 14 TeV for Higgs boson masses near 
$m_H = 160$~GeV.  

With 1 fb$^{-1}$ of integrated luminosity at $\sqrt s = 14$ TeV, using the same analysis cuts as the collaborations, we conclude that a SM Higgs boson with mass $m_H=160$ GeV can be discovered at about the $5\sigma$ level with the ATLAS cuts imposed and about $8 \sigma$ with the CMS cuts imposed.    At $\sqrt s = 10$ TeV, a $3~\sigma$ evidence is possible in the ATLAS case and about 
$5.2~\sigma$ in the CMS case for the same integrated luminosity.  At $\sqrt s = 7$ TeV, the numbers drop to about  $2~\sigma$ in the ATLAS case and about $3.6~\sigma$ in the CMS case. 
 
It is important to bear in mind that our different stated expectations for the two experiments arise from 
the different analysis cuts in the two cases, summarized here in 
Table~\ref{tab: atlas-cms-cut}, particularly the cuts on the dilepton invariant mass $m_{ll}$ 
and on the difference $\Delta\phi_{ll}$ in the azimuthal angles of the two leptons.  
The CMS analysis cuts are more restrictive, tuned more to the region near $160$~GeV, a reason for the greater signal significance and larger statistical uncertainty in our CMS simulations.    

Integrated luminosities of 8 $\rm{fb}^{-1}$ and 3 $\rm{fb}^{-1}$ are needed in the ATLAS case 
at $7$ and $10$~TeV, respectively, for $5\sigma$ level 
discovery of a standard model Higgs boson of mass $m_H = 160$ GeV in the 
$gg\to H\to W^+ W^- \to \ell^+\ell^- + \met$ channel.   In the CMS case, the numbers are 
2 $\rm{fb}^{-1}$ and 1 $\rm{fb}^{-1}$ at $7$ and $10$~TeV, respectively, for $5\sigma$ level 
discovery.    Larger samples would be needed for masses as low as $m_H=140$ GeV or above 
$m_H=180$ GeV.

In the range $140~{\rm GeV}<m_{H}<200~{\rm GeV}$,  
to achieve the same signal sensitivity as at $14$~TeV with $1$~fb$^{-1}$ of integrated 
luminosity, we estimate that  a factor of 6 to 7 more luminosity is required at $7$~TeV 
for the analysis cuts proposed by ATLAS,  and a  factor of about 5 in the CMS case.  
At $10$~TeV, the factor is in the range $\sim 2.5$ for both experiments.   

The acceptances of the Higgs boson signal and dominant backgrounds across the mass range we consider generally increase as the center of mass energy is reduced.  As the cm energy is decreased, the signal cross section is suppressed more than the irreducible background from the $WW$ continuum.  Therefore, more integrated luminosity at a lower cm energy is needed to restore the same significance.  While it is likely that cuts can be tuned to improve the 
expected signal significance at 7~TeV, it also seems likely that Higgs boson discovery
in the $H \to W^+ W^- \to \ell^+ \ell^- + \met $ mode will require more luminosity than currently anticipated. 

Under the assumption that no signal is found, we may restate our results as 95\% exclusion limits on Higgs boson production in $gg$ fusion followed by decay into the $WW$ dilepton mode.  Our results show that with $1~{\rm fb}^{-1}$ of integrated luminosity at 7~TeV, CMS may be able to exclude $m_H$ values of 160 and 180~GeV (and perhaps points in between), while ATLAS may be able to exclude 160~GeV.  Comparison of Figs.~\ref{fig:tevexcludes} and~\ref{fig:LHCexcludes} 
indicates that the Tevatron experiments should remain competitive in the near future provided they achieve analyses based on $10~{\rm fb}^{-1}$ of integrated luminosity.  

\begin{acknowledgments}
The research by E.~L.~B., Q.-H.~C, C.~B.~J, and G.~S. in the High Energy Physics 
Division at Argonne is supported the U.~S.~Department of Energy under
Grant No.~DE-AC02-06CH11357.  Q.-H.~C is also supported in part by the
Argonne National Laboratory and University of Chicago Joint Theory
Institute (JTI) Grant 03921-07-137 and by the U.~S.~Department of
Energy under Grant No.~DE-FG02-90ER40560, and G.~S. is also supported in 
part by the U.~S.~Department of Energy under  Grant No.~DE-FG02-91ER40684.
T.~L. is supported in part by the U.~S.~Department of Energy under Grant 
No.~DE-FG02-90ER40560 and the Fermi-McCormick Fellowship.  
\end{acknowledgments}
\bibliographystyle{apsrev}
\bibliography{reference}

\end{document}